\documentclass{iaus}
\usepackage{graphicx}

\title[The Sulfur Anomaly] 
{The curious conundrum regarding sulfur and oxygen abundances in planetary nebulae }

\author[R.B.C. Henry et al.]   
{R.B.C. Henry$^1$, A. Speck$^2$, A.I. Karakas$^3$,
 \and G.J. Ferland$^4$}

\affiliation{$^1$University of Oklahoma, email: {\tt henry@nhn.ou.edu} \\[\affilskip]
$^2$University of Missouri,  email: {\tt speckan@missouri.edu} \\[\affilskip]
$^3$Mount Stromlo Observatory, email: {\tt akarakas@mso.anu.edu.au} \\[\affilskip]
$^4$University of Kentucky, email: {\tt gjferland@gmail.com}}

\pubyear{2011}
\volume{283}  
\pagerange{}
\jname{Planetary Nebulae: An Eye to the Future}
\editors{A. Manchado \& L. Stanghellini, eds.}
\begin{document}

\maketitle

\begin{abstract}

We carefully consider numerous explanations for the sulfur abundance anomaly in planetary nebulae. No one rationale appears to be satisfactory, and we suggest that the ultimate explanation is likely to be a heretofore unidentified feature of the nebular gas which significantly impacts the sulfur ionization correction factor. 

\end{abstract}

\firstsection 
\section{The Sulfur Anomaly}

\begin{figure}[b]
\begin{center}
 \includegraphics[width=4.0in]{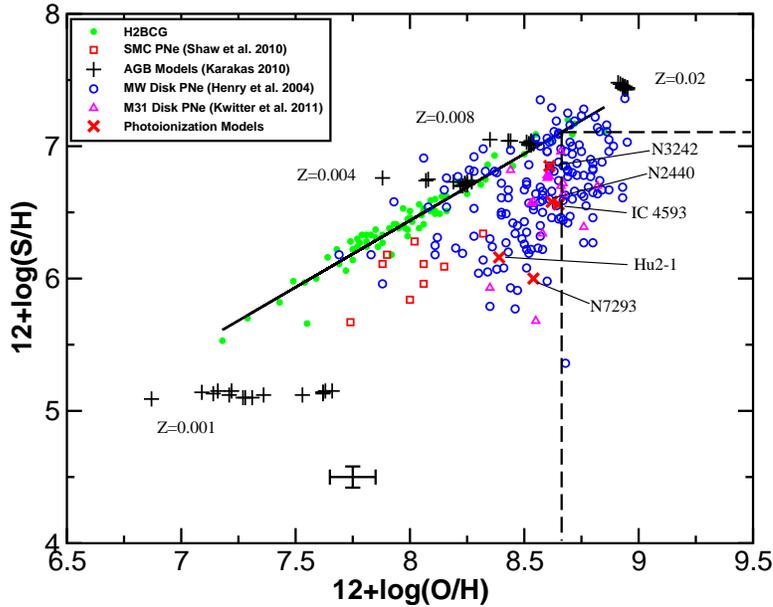} 
 \caption{S versus O for objects in the samples indicated with symbol shape defined in the legend. H2BCG is a composite sample of H II regions and blue compact galaxies compiled from the literature and explained in \cite[Milingo et al. (2010)]{milingo10}. The dashed lines show the solar abundances from \cite[Asplund et al. 2009]{asplund09}.}
   \label{fig1}
\end{center}
\end{figure}

{\it Overview.} The sulfur anomaly (SA) refers to the observation that PN sulfur abundances are systematically lower than those found in most other interstellar abundance probes for the same O abundance (metallicity). See \cite[Henry et al. (2004)]{hbk04}. Fig.~\ref{fig1} shows the SA clearly. The expected relation between S and O is tightly adhered to by a composite sample of H II regions and blue compact galaxies (filled circles; H2BCG). In contrast, PNe in the MWG and M31 disks, and the SMC fall systematically below this track and exhibit much more scatter. Below we summarize the numerous potential explanations for the SA which we have recently explored.


{\it Is The Sulfur Anomaly Related To Other PN Properties?} We quantified the SA by defining the sulfur deficit (SD) as the vertical offset (Fig.~1) of a PN from the H2BCG track. We then compiled values of 19 parameters for each PN (e.g. density, diameter), plotted each against the SD, and computed the correlation coefficient in each case. None of the parameters appeared to be correlated with the SD. 

{\it Is Sulfur Sequestered In Dust Or Molecules?} Sulfur-bearing solids (e.g. MgS) are predicted to form in carbon-rich environments.  Yet we found no correlation between SD and the nebular C/O ratio. Insufficient data currently exist to enable tests for S depletion by molecules.

{\it Stellar Nucleosynthesis of S and O.} Fig. 1 shows (+) the predicted PN abundances of oxygen and sulfur from the AGB models for Z=0.02, 0.008, 0.004 from \cite[Karakas (2010)]{k10} and the Z=0.001 model from \cite[Alves-Brito et al. (2011)]{ab11}. We found insufficient spread in S and O in any of the models and conclude that AGB nucleosynthesis is unable to account for the SA. 

{\it What Other Factors Have Been Tested?} 1. Sulfur abundances derived using direct measurements of S$^{+3}$ ([S IV] 10.5 microns) by \cite[Pottasch \& Bernard-Salas (2006)]{pbs06} in a few cases reduce the SD but clearly do not explain the SA. 2. We have computed an individual photoionization model for each of five PNe representing a broad range in SD. Results shown with Xs and labeled by PN name in Fig. 1 suggest a true S abundance deficiency is responsible for the SA, but this is inconsistent with findings noted above. 3. We tested the effects of reduced and enhanced S$^{+2}$ and S$^{+3}$ dielectronic recombination rates. Such changes had only a minimal effect on the predicted [S III]/[S II] line strength ratios, suggesting that DR is unimportant in explaining the SA.


{\it Conclusions.} We have tested a large range of hypotheses for explaining the SA and have been unsuccessful in identifying a single causal factor. Despite this, we still feel that
the eventual solution will involve the sulfur ICF. A possible solution along this line is offered by \cite[Jacob (2011)]{jacob11}. 

{\it Acknowledgments} We gratefully acknowledge support from the US-NSF (RBCH, GJF) and the NCI National Facility at the Australian National University (AIK). We also thank Matthew Hosak for computing the photoionization models and Karen Kwitter for the use of her PN database.



\begin{thebibliography}{}

\bibitem[Alves-Brito et al. (2011)]{ab11}Alves-Brito, A., Karakas, A., Yong, D., Mel{\'e}ndez, J. \& V{\'a}squez, S. 2011, \textit{A\&A}, in press

\bibitem[Asplund et al. 2009]{asplund09}Asplund, M., Grevesse, N., Sauval, A.J., \& Scott, P. 2009, \textit{ARAA}, 47, 481

\bibitem[Henry et al. (2004)]{hkb04} Henry, R.B.C., Kwitter, K.B. \& Balick, B. 2004, \textit{AJ}, 127, 2284

\bibitem[Jacob 2011]{jacob11}Jacob, R. 2011, \textit{this volume}

\bibitem[Karakas (2010)]{k10}Karakas, A.I., 2010, \textit{MNRAS}, 403, 1413

\bibitem[Kwitter et al. (2011)]{kwitter11}Kwitter, K.B., Lehman, E.M.M., Balick, B., \& Henry, R.B.C. 2011, \textit{in preparation}

\bibitem[Milingo et al. (2010)]{milingo10}Milingo, J.B., Kwitter, K.B., Henry, R.B.C., \& Souza, S.P. 2010, \textit{ApJ}, 711, 619

\bibitem[Shaw et al. (2010]{shaw10}Shaw, R.A., et al. 2010, \textit{ApJ}, 717, 576


\end{thebibliography}
\end{document}